\documentclass[12pt]{article}
\usepackage{a4}
\usepackage{psfig}
\usepackage{amssymb}

\newcommand{\Tr}{\mbox{Tr}\,}

\newcommand{\half}{\frac{1}{2}}
\newcommand{\be}{\begin{equation}}
\newcommand{\ee}{\end{equation}}
\newcommand{\bean}{\begin{eqnarray*}}
\newcommand{\eean}{\end{eqnarray*}}
\newcommand{\bea}{\begin{eqnarray}}
\newcommand{\eea}{\end{eqnarray}}

\newcommand{\bra}{\langle}
\newcommand{\ket}{\rangle}

\newcommand{\vecnul}{{\bf 0}}
\newcommand{\vecp}{{\bf p}}
\newcommand{\veck}{{\bf k}}
\newcommand{\vecx}{{\bf x}}

\newcommand{\bt}{\beta}
\newcommand{\gm}{\gamma}
\newcommand{\dl}{\delta}
\newcommand{\ep}{\epsilon}
\newcommand{\zt}{\zeta}

\newcommand{\lm}{\lambda}

\newcommand{\ph}{\phi}

\newcommand{\om}{\omega}
\newcommand{\Om}{\Omega}

\newcommand{\Gm}{\Gamma}
\newcommand{\Lm}{\Lambda}
\newcommand{\Ph}{\Phi}
\newcommand{\Sg}{\Sigma}
\renewcommand{\theequation}{\arabic{section}.\arabic{equation}}

\begin{document}

\title{
\vskip -100pt
{\begin{normalsize}
\mbox{} \hfill THU-97-15\\
\mbox{} \hfill ITFA-97-24\\
\mbox{} \hfill July 1997\\
\vskip  70pt
\end{normalsize}}
Classical approximation for time dependent quantum field theory: 
diagrammatic analysis for hot scalar fields}
\author{
Gert Aarts$^a$\thanks{email: aarts@fys.ruu.nl}
\addtocounter{footnote}{1}
and
Jan Smit$^{a,b}$\thanks{email: jsmit@phys.uva.nl}\\
\normalsize
{\em $\mbox{}^a$Institute for Theoretical Physics, Utrecht University}\\
\normalsize
{\em            Princetonplein 5, 3508 TA Utrecht, the Netherlands}\\
\normalsize
{\em $\mbox{}^b$Institute for Theoretical Physics, University of Amsterdam}\\
\normalsize  
{\em            Valckenierstraat 65, 1018 XE Amsterdam, the Netherlands}
\normalsize
}
 
\maketitle
 
\renewcommand{\abstractname}{\normalsize Abstract}
\begin{abstract}
\normalsize
We study time dependent correlation functions in hot quantum and 
classical field theory for the $\lm \phi^4$ case. 
We set up the classical analogue of thermal field theory and make a
direct comparison between the quantum and classical diagrams.
A restriction to time independent correlation functions gives the 
connection with conventional
dimensional reduction. If the parameters in the classical 
theory are chosen according to the dimensional reduction matching 
relations, the classical expressions are cutoff independent and
they approximate the quantum expressions,
provided that the external momenta and 
frequencies are small with respect to the temperature. 
\end{abstract}
\newpage

\section{Introduction}

Time dependent phenomena in hot relativistic 
quantum field theory play an important 
role in cosmology (baryogenesis, inflation) and heavy ion collisions. 
Non-perturbative calculations of these phenomena under general circumstances
are very difficult. 
A potentially powerful approach is the classical approximation, in 
combination with numerical simulations.
When this approach was introduced for the computation of the
rate of sphaleron transitions \cite{GrRu88,GrRuSh89},
the idea was to choose the cutoff of the order of the temperature,
since only the low momentum modes of the fields are
expected to behave classically at high temperature.
This raises the question how the results of the simulations depend on 
the cutoff.
Improving the method by consistently taking the high momentum 
modes into account (by integrating them out of the original quantum theory
\cite{TaSm95,BoMcLeSm95}) leads to very complicated effective equations 
of motions \cite{GrMu96,HuSo97}.
 
To try to get around these problems we investigated in \cite{AaSm97} the effect
of taking the momentum cutoff to infinity, 
in hot time dependent {\em classical} $\lm\phi^4$ theory. 
It turned out that the cutoff 
dependence could simply be canceled with local counterterms, 
as in the dimensional reduction approach for time independent quantities. 
Furthermore we found that with a suitable choice of the parameters 
in the classical hamiltonian, the classical plasmon damping rate 
is the leading order (in $\lambda$) result of the quantum theory.
This result has been confirmed in a recent calculation \cite{BuJa97}.

In this paper we give a 
diagrammatic comparision between the real-time correlation 
functions in the quantum theory and the classical theory for $\lambda 
\phi^4$. We show that there is a direct correspondence between 
quantum and classical diagrams,
and that the classical result is the leading order 
quantum result at high $T$, provided that
the classical parameters are chosen correctly, 
and the theory is weakly coupled.

After a short review of the dimensional reduction approach for time 
independent correlation functions, we introduce a version of the real-time 
formalism for finite temperature quantum field theory which is 
appropriate for a direct comparision with the classical theory. 
We then introduce the classical approximation and show how to identify 
diagrams that arise in perturbation theory. After this setup, 
we calculate the two point function to two loops and the four point 
function to one loop, and compare the quantum with the classical results.
Throughout we work with $\hbar = 1$, except when we explicitly want to 
indicate what part of the results is really classical and what part 
is quantum.

\section{Dimensional reduction for time independent quantities}
\label{sectiondimred}
We start with a brief summary of the dimensional reduction approach for 
time independent quantities, see for example 
\cite{Faea94, Kaea96a, Ja96}.

Static quantities in a quantum field theory at finite temperature are 
most easily calculated in the imaginary time or Matsubara formalism 
\cite{Kap}. The partition function is given by
\[ Z = \int {\cal D}\phi\, e^{-S_E},\]
with the euclidean action
\be
\label{eqeucl}
S_E = \int_0^\beta d\tau \left[\int d^3x 
\left(\half (\partial_\tau \phi)^2 + {\rm ct}\right)
+ \bar V(\phi)\right],
\ee
where $\bar V$ is the potential energy
\[
\bar V(\phi) = \int d^3x \left(\half (\nabla \phi )^2 + \half \bar m^2 \phi^2 +
\frac{\bar\lambda}{4!} \phi^4\right),
\]
and ${\rm ct}$ denotes the counterterms.
We assume dimensional renormalization in the $\overline{\mbox{MS}}$ scheme and 
$\bar m$ and $\bar\lm$ are the corresponding parameters. The
counterterms are given in appendix \ref{appendixcounter}.
The field satisfies periodic boundary conditions 
$\phi(\vecx, 0) = \phi(\vecx, \beta)$, 
which leads to the decomposition 
\[  
\phi(\vecx, \tau) = T\sum_n e^{i\om_n\tau}\phi_n(\vecx),
\]
where $\om_n = 2\pi nT$ are the Matsubara frequencies. In the dimensional 
reduction approach, an effective theory is constructed for
$\phi_{0}$. This effective theory is chosen to have the same form as 
the $\phi_{0}$ part of the full quantum theory, but with effective 
parameters. Hence the partition function for the dimensionally reduced 
theory is given by 
\be
\label{eqpartfunc}
 Z_{DR} = \int {\cal D}\phi_{0}\, e^{-\beta V},
\ee
with 
\[
V = \int d^3x \left( \half (\nabla \phi_0)^2 
+ \half \nu^2\phi_0^2 
+ \frac{\lambda}{4!}\phi_0^4 + \ep\right).
\]
We did not absorb the prefactor $\beta$ in the field and coupling 
constant as is usually done. For completeness we included 
an effective cosmological constant $\ep$. 

The parameters in the effective theory are determined by perturbatively 
matching correlation functions in the effective theory to 
correlation functions in the full theory.
The effective theory needs regularization and we use a momentum cutoff 
$\Lambda$. Because it is superrenormalizable, only the effective mass 
parameter and the cosmological constant receive a $\Lambda$ dependent part. 
The divergent integrals appearing in the scalar field selfenergy 
corresponding to the one loop leaf diagram and the two loop setting sun 
diagram are

\bea
\nonumber
\frac{\lambda T}{2} \int_{|\veck| < \Lm} 
\frac{d^3k}{(2\pi)^3} \frac{1}{\om_\veck^2} 
&=& 
\lambda T \left( \frac{\Lambda }{4\pi^2} - \frac{m}{8\pi} \right) ,\\ 
-\frac{\lm^2 T^2}{6} 
\int_{|\veck_{1,2,3} < \Lm|} d\Ph_{123}(\vecnul)
\frac{8}{\om_{\veck_1}\om_{\veck_2}\om_{\veck_3}
}
&=& 
\label{statsetsun}
-\lambda^2T^2 \left(
\frac{1}{16\pi^2}\log\frac{\Lambda}{3m} + L_0
\right),
\eea
where
\bea
\om_{\veck} &=& \sqrt{ \veck^2 + m^2}\\
\label{eqdPh}
 d\Ph_{123}(\vecp) &=& 
\frac{d^3k_1}{(2\pi)^32\om_{\veck_1}}\,
\frac{d^3k_2}{(2\pi)^32\om_{\veck_2}}\,
\frac{d^3k_3}{(2\pi)^32\om_{\veck_3}}\,
(2\pi)^3 \dl(\vecp - \veck_1 - \veck_2 - \veck_3).
\eea
The regularization dependent constant $L_0 = 0.0067$ 
\cite{Ja96}. 
The result of matching the two point function \cite{Faea94} to two loops 
can be written in the form
\be
\label{eqmatch1}
\nu^2 = m^2 -\delta m^2, \;\;\; \;
\delta m^2 =  \lambda m_1^2 + \lambda^2 m_2^2,
\ee
with\footnote{The choice here of the finite
parts of $m_{1,2}^2$ differs from that in ref.\ \cite{AaSm97}.}
\bea 
\label{eqmatchm}
m^2 &=&  \bar m^2 + m_{\rm th}^2,\;\;\;\;
m_{\rm th}^2 = \frac{\bar\lambda T^2}{24}, \\ 
\label{ct1} 
m_1^2 &=&  \frac{\Lambda T}{4\pi^2} + \frac{m^2}{32\pi^2}\log 
\frac{\bar\mu^2}{\mu_T^2},
\;\;\;\;
\mu_T \equiv 4\pi\exp(-\gm_E) T,\\ 
 m_2^2 &=& -T^2\left( \frac{1}{16\pi^2}\left[ \log \frac{\Lm}{3T} - c - \half
\right] + L_0\right),
\label{ct2}
\eea
Here $\bar\mu$ is the scale parameter in the $\overline{\mbox{MS}}$ scheme and 
the constant $c= -0.34$ is given 
in \cite{Faea94, Kaea96a}. 
For future convenience we have separated the effective mass 
parameter $\nu^2$ into a tree level mass $m^2$, and two counterterms 
$\lambda m_1^2$ and $\lambda^2 m_2^2$. 
The matching (\ref{ct2}) is such that the three and four dimensional
renormalized selfenergies coincide at zero momentum.

For the coupling constant to one loop, the relation is 
\be
\label{eqmatchl}
\lambda = \bar\lambda - \frac{3\bar\lambda^2}{32\pi^2} \log 
\frac{\bar\mu^2}{\mu_T^2}.
\ee

Finally for the cosmological constant, matched to one loop, we get 
\be
\label{eqmatch2}
\ep =  \bar\ep - \frac{\pi^2 T^4}{90} + \frac{m^2 T^2}{24}
-  \frac{ m^4}{64\pi^2} \log \frac{\bar\mu^2}{\mu_T^2}
- \frac{\Lambda^3 T}{6\pi^2}\left(\ln\frac{\Lm}{T} - \frac{1}{3}\right) + 
\frac{\Lm Tm^2}{4\pi^2}.
\ee
 Note that the Rayleigh-Jeans divergence in the energy 
density is canceled, and the Stefan-Boltzmann law is put in by hand 
through the matching procedure.

Corrections to the dimensional reduction approximation are 
small when the thermal mass and external momenta are small compared to the 
temperature, i.e.\ 
\[
\frac{m_{\rm th}^2}{T^2} =  \frac{\bar\lambda}{24} \ll 1,
\;\;\;\; \frac{p^2}{T^2} \ll 1.
\]
This is satisfied when the theory is weakly coupled and physical
observables are dominated by momenta smaller than ${\cal O}(\sqrt{\lm}T)$.  

This completes the standard description of the 
dimensional reduction approach for time independent quantities. In the 
following sections we move on to time dependent correlation functions.

\section{Real time formulation of quantum field theory at finite 
temperature}
\setcounter{equation}{0}
In the preceding section we were only interested in static quantities. 
For this the imaginary time formalism is very useful. However, to make a 
direct diagrammatic analysis of time dependent correlation functions, a 
more suitable formulation is the real-time formalism \cite{LavW, LeB}.
In this section we review 
a version of the real-time formulation that is 
very convenient for a comparison with the classical theory.
\begin{figure}
\centerline{\psfig{figure=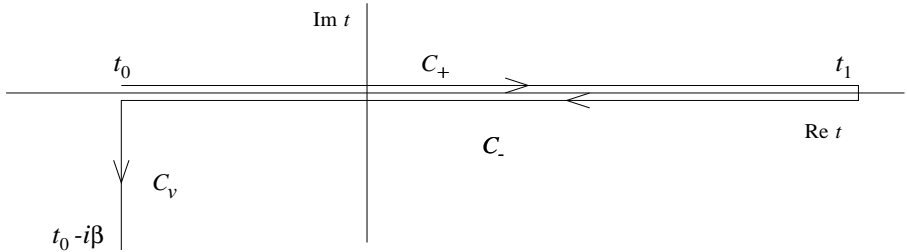,width=12cm}}
\caption{Keldyshcontour in the complex $t$-plane.}
\label{figkeldysh}
\end{figure}

The real-time expectation value
\be
\label{eqdefobs}
\bra O(t) \ket = Z^{-1}\Tr e^{-\bt H}\, O(t),
\ee
with
\[ Z = \Tr e^{-\bt H}, \;\;\;\; O(t) = e^{iHt}\, O e^{-iHt},
\]
can be represented by the path integral
\[
\bra O(t) \ket = Z^{-1} \int {\cal D}\phi \, e^{iS} O(t),
\]
with the action\footnote{We use a space favoured metric (-1,1,1,1); 
note in particular that $k^0 = -k_0$ corresponds to $k^0$ in the
time favoured metric (1,-1,-1,-1).}
\be
\label{eqaction}
 S = -\int_C dt
\int d^3x \left(\half \partial_\mu \phi \partial^\mu \phi + 
\half \bar m^2 \phi^2 +\frac{\bar\lambda}{4!}\phi^4
+ {\rm ct}\right).
\ee
The subscript $C$ labels the contour\footnote{Other contours are also 
possible \cite{LavW}, but we prefer to use the Keldysh contour.} 
in the complex $t$-plane shown in fig \ref{figkeldysh}. 
The result is obviously independent of the initial and final 
times $t_0$ and $t_1$. In thermal field theory one lets $t_0 \to -\infty$
and $t_1 \to +\infty$. This leads to the introduction of two fields
$\ph_+$ and $\ph_-$ which live on the $+$ and $-$ branches of the contour.
It has been shown \cite{Nie89, EvPe95} 
that the field on the vertical branch of the contour at $t_0 = -\infty$
can be ignored for the calculation of expectation values 
as in (\ref{eqdefobs}),
such that in perturbation theory only the propagators and vertex
functions for the $\pm$ field combinations enter.
Explicitly, the vertex functions correspond to the interaction
\[
-\int_{-\infty}^{\infty} dt \int d^3x\, \frac{\bar\lm}{4!}\,(\ph_+^4 - \ph_-^4). 
\]
However, the form of the propagator for the $\pm$ fields reflects the
vertical branch of the contour at $t_0 = -\infty$. For example, 
the full propagator for the $\pm$ fields
is still given by the original operator representation
($t=x^0$),
\bean
G^{++}(x-x') &=& \theta(t-t')G^>(x-x') +   \theta(t'-t)G^<(x-x'),\\
G^{+-}(x-x') &=& G^<(x-x'),\;\;\;\;\;
G^{-+}(x-x') = G^>(x-x'),\\
G^{--}(x-x') &=& \theta(t'-t)G^>(x-x') +   \theta(t-t')G^<(x-x'),
\eean
with 
\bean 
G^>(x-x') &=& i\bra \phi(x)\phi(x')\ket,\;\;\;\;\;\;
G^<(x-x') = i\bra \phi(x')\phi(x)\ket,
\eean
These expectation values are related by the well known 
KMS condition
\be
\label{eqkms}
 G^>(\vecx - \vecx',t-t') = G^<(\vecx-\vecx',t-t'+i\beta),
\ee
which follows from the cyclicity of the trace and the fact that the time 
evolution and the canonical average are governed by the same hamiltonian. 
Using the KMS condition to determine the explicit form of 
the propagators takes into account the vertical branch \cite{Nie89, EvPe95}. 
We stress this point because we will follow the same steps when we consider 
the classical theory. 

To make a direct relation with classical correlation functions, 
it is convenient to go to another basis, given by
\be
\label{eqrotphi}
 \left( \begin{array}{c} \phi_1\\
\phi_2
\end{array}\right) = 
\left(\begin{array}{c}
(\phi_+ +\phi_-)/2
\\
\phi_+ - \phi_-
\end{array}\right) =
R \left( \begin{array}{c}
\phi_+\\ 
\phi_-
\end{array}\right) 
, 
\ee 
with 
\[ R = \left(
\begin{array}{cc} 1/2 & 1/2 \\ 
1 & -1  
\end{array}\right).\] 
This is a variation of the Keldysh basis, where $R$ is 
a rotation over $\pi/4$. In fact, the basis we use here is closely 
connected to the influence functional approach of Feynman and Vernon 
(for a recent application, see e.g. \cite{GrMu96} where semiclassical 
equations are derived for $\phi_1$).

In matrix form,
\[
{\bf G}(x-x') = 
\left( \begin{array}{cc} 
G^{++}(x-x') & G^{+-}(x-x')\\ 
G^{-+}(x-x') & G^{--}(x-x')
\end{array}\right),
\]
the propagator is now given by
\be 
\label{eqpropkel}
{\bf G}(x-x') \to R {\bf G}(x-x')R^T = 
\left( \begin{array}{cc} iF(x-x') & G^R(x-x')\\
G^A(x-x') & 0 
\end{array}\right), 
\ee
with 
\bea
F(x-x') &=& 
\half \bra \phi(x)\phi(x')+\phi(x')\phi(x)\ket ,\\
\label{eqgrgaqm}
G^R(x-x') &=& G^A(x'-x) = i\theta(t-t')\bra[\phi(x),\phi(x')]\ket.
\eea 
The last equation can be used to define the spectral function\footnote{We have
chosen the convention that $G^R(x), G^A(x), F(x)$ and $\rho(x)$ are all 
real.} \be
\label{eqspfuqm}
 \rho(x-x') = i\bra[\phi(x),\phi(x')]\ket 
= G^R(x-x')-G^A(x-x'). 
\ee

We will now use the KMS condition to relate the various introduced quantities.
After Fourier transformation to momentum space, according to 
($k = (k^0,\veck)$)
\[ G^>(k) = \int d^4x\, e^{ik^0t-i\veck\cdot\vecx}\,G^>(x),\]
the KMS condition (\ref{eqkms}) reads
\[ G^>(k) = e^{\beta k^0}\,G^<(k).\] 
{}From the KMS condition we find the following expressions in  
terms of the spectral function
\bea
\nonumber 
G^>(k) &=& [n(k^0) + 1] \rho(k),\\
\nonumber 
G^<(k) &=& n(k^0) \rho(k),\\
\label{eqkmsqm}
F(k) &=& -i[n(k^0)+\half]\rho(k),\\
\nonumber
n(k^0) &=& \frac{1}{e^{\bt k^0} -1}.
\eea
For $k^0 >0$, $n(k^0)$ is the Bose distribution.

Notice that all components of the propagator are determined by the 
retarded
propagator $G^R$. We could have used (\ref{eqkmsqm}) to reduce the 
matrix propagator to a diagonal matrix $\mbox{diag}(G^R, G^A)$, as is 
done in the so-called R/A formulation \cite{AuBe92}.\footnote{Note, however, 
that also in the R/A formulation the $F$ propagator will appear;
it is $\phi(k)$ in \cite{AuBe92}.} However, the formulation we use makes the 
comparison with the classical theory more direct, as we will show.

We end this section with some useful properties of the self energy 
\be  
\label{eqnselfenergy}
{\bf \Sigma}(k) = {\bf G}^{-1}(k) - {\bf G}_0^{-1}(k), 
\ee 
with ${\bf G}$ given by the 1,2-form (\ref{eqpropkel}) and the subscript $0$ 
indicates the free propagator. The inverse propagator is given by
\[
{\bf G}^{-1}(k) = 
\left(\begin{array}{cc}
0 & G^A(k)^{-1} \\ G^R(k)^{-1} & [n(k^0) + \half][G^{R}(k)^{-1} - G^A(k)^{-1}]
\end{array}\right),
\]
where we used (\ref{eqkmsqm}) and (\ref{eqspfuqm}). It follows that $\Sg$ has the form
\[
{\bf \Sigma}(k) = \left( \begin{array}{cc} 
\Sigma_{11}(k) & \Sigma_{12}(k)\\
\Sigma_{21}(k) & \Sigma_{22}(k) \end{array}
\right)  =
 \left( \begin{array}{cc} 
0 & \Sigma_{A}(k)\\
\Sigma_{R}(k) & i\Sigma_F(k) \end{array}
\right),
\]
with
\bea
\label{eqpi1}
\Sg_R(k) &=& \Sg_A(-k) = \Sg_A^*(k) = G^{R}(k)^{-1} - G_0^R(k)^{-1},\\
\label{eqpiF}
\Sg_F(k) &=& -i[n(k^0) + \half] [\Sg_R(k) - \Sg_A(k)] 
\eea
{}From this we conclude that a calculation of the retarded 
self energy $\Sigma_R(k)$ is sufficient to determine also the other components.

\section{Perturbation theory in the quantum theory}
\setcounter{equation}{0}
\label{sectionpertqm}
Due to the fact that $\phi(x)$ and $\phi(x')$ do not 
commute in a quantum theory, different orderings of the fields along the 
contour give different $n$ point functions. 
In the classical theory, however, the typical two point function 
to be calculated is
\be
\label{eqclascor}
\bra \phi(x)\phi(x')\ket_{\rm cl},
\ee
with no special ordering (classically the fields commute of course). The 
quantum  two point function we are interested in, is therefore
\[ 
\half\bra \phi(x)\phi(x')+\phi(x')\phi(x)  \ket = F(x-x') = \bra 
\phi_1(x)\phi_1(x')\ket,
\]
which reduces formally to (\ref{eqclascor}) in the limit $\hbar \to 0$.
We consider  $n$ point connected Green functions where the 
external field is $\phi_1 = \half (\phi_++\phi_-)$.

The free propagators are given by
\bea
\label{eqGrGa} G^R_0(k) &=& G_0^A(-k) = {G_0^A}^*(k) = 
\frac{1}{\om_\veck^2-(k^0+i\ep)^2},\\
\label{eqkmsF} F_0(k) &=& -i(n(k^0)+\half)\left(G_0^R(k) -G_0^A(k)\right) \\
\nonumber
&\stackrel{\ep \downarrow 0}{=}&
(n(k^0)+\half)\ep(k^0)2\pi\delta(k_0^2-\om_\veck^2), 
\eea
with 
$\ep(k^0) = \theta(k^0)-\theta(-k^0)$.
Note that the $i\ep$ dependence is unambiguously 
determined by the $i\ep$ dependence of the retarded (and advanced) Green 
functions. We 
keep $\ep$ finite throughout, to avoid products of delta functions.

The interaction part has to be expressed in terms of the $\phi_1, \phi_2$ 
fields using (\ref{eqrotphi}), which gives 
\be
\nonumber
-\int_{-\infty}^{\infty}dt \int d^3x \,
\frac{\bar\lambda}{4!}\left(\phi_+^4 -\phi_-^4\right)
=
\label{eqpotrot} -\int_{-\infty}^\infty dt \int d^3x 
\,\frac{\bar\lambda}{4!} 
\left(4 \phi_1^3 \phi_2  + \phi_1 \phi_2^3 \right).
\ee
\begin{figure}
\centerline{\psfig{figure=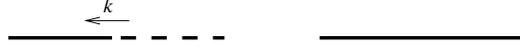,width=7cm}}
\caption{Propagators, a)  $G_0^R(k) = G_0^A(-k)$, b) $iF_0(k)$.}
\label{figprops} 
\end{figure}
\begin{figure}
\vspace{1cm}
\centerline{\psfig{figure=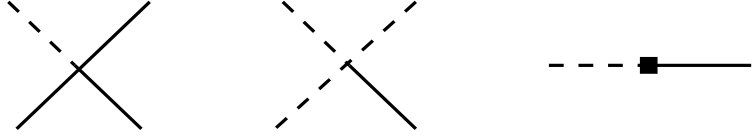,width=10cm}}
\caption{Vertices, 
a) $-\bar\lambda$, 
b) $-\bar\lambda/4$,
c) $m^2_{\rm th}$.} 
\label{figvertices} 
\end{figure}
The Feynman rules are given in figs \ref{figprops} and 
\ref{figvertices}a, b. 
We represent the $\phi_1$ field with a full line, 
and the $\phi_2$ field with a dashed line. Hence the retarded and advanced Green 
functions, 
which interpolate between a $\phi_1$ and a $\phi_2$ field, are indicated with 
a dashed-full line.

The diagrams we are about to calculate contain the normal four 
dimensional, zero temperature divergences. 
As 
mentioned before, these will be treated using dimensional 
renormalization in the $\overline{\mbox{MS}}$ scheme. 
The counterterms can be found in appendix \ref{appendixcounter}.

As is well known, a consistent perturbative treatment needs a resummation 
of the so-called hard thermal loops (HTL's). In the scalar theory, 
the essential HTL is only the leaf diagram, which gives rise to 
the thermal mass $m^2_{\rm th} = \bar\lambda T^2/24$. Resummation 
can be done by 
a simple  addition  and subtraction of the thermal mass
\cite{BaMa91, Pa92, WaHe96}. The mass parameter that enters in the 
free propagator is now  $m^2 = \bar m^2 + m^2_{\rm th}$, furthermore there 
is  a new vertex, as shown in fig \ref{figvertices}c.  We will not 
try to improve perturbation theory 
any further, because our  interest lies in the comparison with the 
classical theory, not in the improvement of perturbation theory.

We are now ready to calculate correlation functions in the quantum theory, but 
because we want to compare these  with  the ones obtained in the classical 
approximation, we first discuss the classical theory.

\section{Classical correlation functions}
\setcounter{equation}{0}
In the following section we consider correlation functions in 
the classical theory. 
After the definition of classical correlation functions, 
we discuss the choice of 
parameters in the (effective) hamiltonian. Since the classical theory 
is supposed to be 
an approximation to the quantum theory at high temperature, the 
parameters in the 
classical hamiltonian are not arbitrary. We determine the parameters by 
restricting ourselves to equal time correlation functions, and then  
using the connection with the conventional dimensional reduction 
approach, as described in section \ref{sectiondimred}.
We test whether this procedure 
makes sense for time dependent correlation functions as well by a 
calculation of the 
classical two point function up to two loops  and the classical  four 
point function up to 
one loop and comparing these with the high temperature 
expressions from the quantum theory.

We start by writing the hamiltonian, that corresponds to the classical 
action (\ref{eqaction}), 
\[  
H = \int d^3x \left(\half \pi^2 + \half (\nabla \phi)^2 + \half \nu^2
\phi^2 + \frac{1}{4!}\lambda\phi^4 + \epsilon \right).
\]   
The classical two point function is defined in the following way 
\be
\label{eqdef2punt}
S(x-x') \equiv \bra \phi(x)\phi(x')\ket_{\rm cl} \equiv \frac{
\int  {\cal D}\pi {\cal D}\phi\, e^{-\beta H(\pi,\phi)}\phi(x)\phi(x')}{
\int  {\cal D}\pi {\cal D}\phi\, e^{-\beta H(\pi, \phi)}},
\ee
with $\phi(x)$ the solution of the equations of motion
\bea
\nonumber
\dot{\phi}(x) &=& \{\phi(x), H\},\\
\label{eqmotion}
\dot{\pi}(x) &=& \{\pi(x), H\},
\eea
the curly brackets denote the Poisson brackets.
The initial conditions of these equations of motions are 
\[ \phi(\vecx, t_0) = \phi(\vecx),\;\;\;\; \pi(\vecx, t_0) = \pi(\vecx).\]
The integration in (\ref{eqdef2punt}) is over the initial conditions, 
weighted with the Boltmann weight.

As explained in the beginning of this section, the 
parameters in the hamiltonian are not arbitrary. We determine them  by 
studying static correlation functions.  
If we restrict ourselves to $t = t'$,
the fields are related by a canonical transformation with those at the initial
time $t_0$. Then the integration over the canonical momenta is trivial, 
and we find the following expression 
\be
S(\vecx-\vecx', 0) = \bra \phi(\vecx,t)\phi(\vecx',t)\ket_{\rm cl} = \frac{
\int {\cal D}\phi\, e^{-\beta V(\phi)}\phi(\vecx)\phi(\vecx')}{
\int {\cal D}\phi\, e^{-\beta V(\phi)}},
\ee
which is just a correlation function in a three dimensional field theory. In 
fact, the partition function is similar in form to the one obtained in 
the dimensional reduction approach (\ref{eqpartfunc}). Using this 
observation we take the parameters in the effective hamiltonian to be 
given by the dimensional reduction matching relations, i.e. 
(\ref{eqmatch1})-(\ref{eqmatch2}). The only thing that is not determined 
this way is a possible prefactor in front of the kinetic part of the 
hamiltonian, i.e. 
$ \pi^2 \to  z_\pi\pi^2$. 
We shall find that $z_\pi = 1$ to leading order in the matching of the classical
and quantum theories.

Note that although we only match time 
independent quantities, this will be sufficient to make the time 
dependent quantities well-defined and useful (in \cite{AaSm97} we studied 
in particular the finiteness of the two point function). 

Now that we have determined the parameters in the hamiltonian, we can 
discuss the properties of time dependent correlation functions.
To make the comparison with the quantum theory straightforward, we 
introduce the 
classical counterparts of (\ref{eqgrgaqm}) and (\ref{eqspfuqm})
\be 
\label{eqgrgacl}
G^R_{\rm cl}(x-x') = G^A_{\rm cl}(x'-x) = - \theta(t-t')\bra 
\{\phi(x),\phi(x')\}\ket_{\rm cl}, \ee
\be
\rho_{\rm cl}(x-x') =  -\bra \{\phi(x),\phi(x')\}\ket_{\rm cl} = 
G^R_{\rm cl}(x-x')-G^A_{\rm cl}(x-x'), \ee
and the classical KMS condition (see appendix \ref{appendixKMS})
\be
\label{eqkmsclassical}
\frac{d}{dt}S(x-x') = - T\rho_{cl}(x-x').
\ee

If we specialize to the unperturbed problem ($\lambda = 0$), we can immediately 
solve the equations of motion and find
\be
\label{eqfree}
\phi_0(\veck,t) = \int d^3x\, e^{-i\veck\cdot\vecx}\phi_0(\vecx, t) =
\phi(\veck)\cos \omega_\veck(t-t_0) +
\frac{\pi(\veck)}{\omega_\veck}\sin \omega_\veck(t-t_0),
\ee
with $\omega_\veck^2 = \veck^2+m^2$, and where $\phi(\veck)$ and $\pi(\veck)$ are 
the Fourier components 
of the initial conditions. It is now easy to calculate the Poisson brackets
in (\ref{eqgrgacl}) to find
\be 
\label{eqgrclt}
G^R_0(\veck, t-t') = \theta(t-t')\frac{\sin \om_\veck (t-t')}{\om_\veck}.
\ee
Using the KMS condition (\ref{eqkmsclassical})
we find for the free two point function (up to a constant)
\be 
\label{eqs0}
S_0(\veck, t-t') =   T\frac{\cos \om_\veck (t-t')}{\om_\veck^2},
\ee
which can also be found by explicitly performing the Gaussian 
integrations in (\ref{eqdef2punt}).
It is easy to make a connection with the quantum theory at this point.
Fourier transforming the free retarded Green function in the quantum 
theory  (\ref{eqGrGa}) 
back to time space  gives (\ref{eqgrclt}). $G_0^R(x-x')$ is independent 
of $T$ and $\hbar$.  Hence we drop the subscript `cl' for the free retarded and advanced 
Green functions.
Fourier transforming the thermal propagator in the quantum theory, 
(\ref{eqkmsF}), gives 
\be
\label{eqfqmt}
F_0(\veck, t-t') = [n(\om_\veck)+\half]\frac{\cos \om_\veck 
(t-t')}{\om_\veck} = 
T\sum_n \frac{\cos \om_\veck (t-t')}{\om_n^2+\om_\veck^2},\ee
with the Matsubara frequenties $\om_n = 2\pi nT$.
We have used the identity
\[
\frac{1}{\om_\veck}[n(\om_\veck)+\half] =T\sum_{n=-\infty}^{\infty} 
\frac{1}{\om_n^2+\om_\veck^2},\]
to obtain the last equality.
Note that we have written the real-time correlation function as a  
sum over the Matsubara modes, familiar from the euclidean formulation. 
The $n=0$ mode in (\ref{eqfqmt}) is precisely the classical propagator 
(\ref{eqs0}).

We conclude this section with a short discussion of $\hbar$. 
Up to now we have put $\hbar=c =1$. However, $\hbar$ can always be 
restored by 
dimensional analysis, keeping $c=1$. Since it has the dimension of energy 
times length ($[\hbar] = [E\, l]$), we no longer have $[E] = [l^{-1}]$.
By inspection of the classical hamiltonian, we find the following dimensions
$[\phi^2] = [E\, l^{-1}]$, $[m] = [l^{-1}]$, 
$[\lm] = [E^{-1}\, l^{-1}]$; furthermore
$[\vecp] = [p^0] = [l^{-1}]$ and for the selfenergy $[\Sg] = [l^{-2}]$.
Using this we find that the thermal mass is proportional to $\hbar^{-1/2}$, 
$m^2_{\rm th} = \bar\lambda T^2/24\hbar$. 
Thus we find that $\hbar$ is introduced in the classical theory
through the matching relations. 

In the following section we discuss the perturbation expansion in the 
classical theory.

\section{Perturbation theory in the classical theory}
\setcounter{equation}{0}
\label{sectionrulescl}

To study classical time dependent correlation functions in 
perturbation theory we can 
follow two approaches. The first approach is the one we followed 
in \cite{AaSm97}, the 
averaging over the initial conditions is done at $t = t_0$, and $t_0$ 
is kept finite. 
Both the solution to the equations of motion 
and the Boltmann weight are expanded in $\lambda$ and correlation functions are 
calculated up to the desired order. As we have shown in \cite{AaSm97} 
the final answer 
is independent of $t_0$, as expected in equilibrium. 

We will now follow the thermal field theory approach, i.e. we use the 
classical KMS 
condition (\ref{eqkmsclassical}) to determine $S_0(k)$ to be 
\be 
S_0(k) = S_0(\veck, k^0) = -i\frac{T}{k^0}\left( G_0^R(k) - G_0^A(k)\right),
\ee
which is the analogue of (\ref{eqkmsqm}). 
The Boltzmann weight $\exp(-\bt H)$
does not enter the perturbative calculations anymore 
and throughout we work in temporal momentum space. In appendix
\ref{appendixmassderivative} 
we 
illustrate both approaches with a simple example.

To calculate correlation functions in perturbation theory, we need to solve the 
equations of motion (\ref{eqmotion}) perturbatively. 
In terms of the solution to the unperturbed problem $\phi_0(x)$ and the 
retarded Green function $G_0^R(x)$, the 
solution of the equations of motion to second order in the coupling constant 
given by	
 \bea 
\nonumber
\phi(x) &=& \phi_0(x) + \lambda \phi_1(x) + \lambda^2 \phi_2(x)\\
\nonumber
 &=& 
\phi_0(x) 
 - \lambda \int d^4x'\, G^R_0(x-x')\frac{1}{3!}\phi_0^3(x')\\ 
\label{eqsol2order}
&&
\mbox{} + \lambda^2 \int d^4x'\,  G^R_0(x-x')\frac{1}{2!}\phi_0^2(x') 
\int d^4x''\, G^R_0(x'-x'')\frac{1}{3!}\phi_0^3(x''). 
\eea

If we now want to calculate correlation functions like 
$\bra\phi(x_1)\phi(x_2)\ket_{\rm cl}$ up to a certain 
order, we expand the 
$\phi(x)$'s as in (\ref{eqsol2order}), and order terms according to 
their power of $\lambda$. This gives rise to diagrams in the following way. 
The vertices are coming from the integrations over the retarded Green 
functions and the zero order solutions, as in (\ref{eqsol2order}). The lines in 
the diagrams are either $G_0^R$ lines, coming from the solution 
(\ref{eqsol2order}), or $S_0$ lines, coming from the contractions of products 
of $\phi_0$. 
In diagrams we use the same representation as in the quantum case, i.e. 
the dashed-full lines represent the retarded and advanced Green 
functions, and the full line the thermal propagator $S_0$.

A straightforward calculation of correlation functions, as 
defined in (\ref{eqdef2punt}), results in the connected diagrams. 
As can be deduced 
from the expressions in \cite{AaSm97}, the two point function 
can be written in the form
\[ S = S_0 - S_0\Sigma_{A, \rm cl}G^A - G_0^R\Sigma_{R, \rm cl}S - 
G_0^R\Sigma_{F, \rm cl}G^A,\] 
which is just the 
analogue of the $F$ component of 
${\bf G} = {\bf G}_0 - {\bf G_0} {\bf \Sg} {\bf G}$
obtained by multiplication of (\ref{eqnselfenergy}) 
with ${\bf G}_0$ and ${\bf G}$ from the left and right.
In the following we will  
truncate the external lines and concern ourselves only with the 1PI diagrams. 
We use the same $(1,2)$-notation as in the quantum case to denote the 
different types of diagrams, e.g. the classical retarded self energy is 
written as $\Sigma_{R, \rm cl}(p) = \Sigma_{21, \rm cl}(p)$.

A last remark is related to the resummation in the quantum 
theory. In the classical perturbation expansion, the thermal mass which 
is generated through the matching prescription (\ref{eqmatchm}), is 
included in the zero 
order mass parameter, just as in the quantum theory after resummation.

We are now ready to compute time dependent  correlation functions and 
compare the classical result with the full quantum field theory result. 
We will study the self energy and the four point function in the 
following sections.

\section{The self energy}
\label{sectionqmselfenergy}
In this section we calculate the self energy to two loops in both the quantum 
and the classical theory. 
To simplify the notation in the quantum theory and avoid writing $\int
d^{4-2\ep} k$ etc, we shall assume the counterterms of the
$\overline{\mbox{MS}}$ scheme to be absorbed into the symbol $\int d^4
k$.

\subsubsection*{First order self energy}

\begin{figure}
\centerline{\psfig{figure=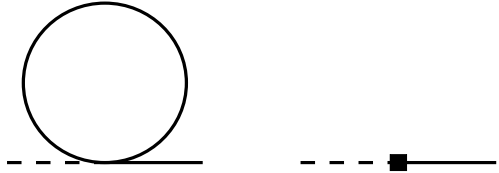,width=6.6cm}}
\caption{Retarded self energy $\Sigma_{R}^{(1)}$; the second diagram,
coming from resummation, is only present in the quantum theory.}
\label{figtwofirst}
\end{figure}

For the one loop retarded self energy in the quantum theory, we get
\be \Sigma_{R}^{(1)} = \half\bar\lambda \int \frac{d^4
k}{(2\pi)^4}F_0(k) - m_{\rm th}^2.  \ee The diagrams are shown in fig
\ref{figtwofirst}.  Performing the integral leads to
\[
\Sigma_{R}^{(1)} = - \bar\lambda\frac{mT}{8\pi}
-\bar\lambda\frac{m^2}{32\pi^2}\log \frac{\bar\mu^2}{\mu^2_T} + {\cal
O}(T^{-2}).
\]
Furthermore we find from (\ref{eqpi1}) $\Sigma_{A}^{(1)} =
\Sigma_{R}^{(1)}$ and hence from (\ref{eqpiF}) $\Sigma_{F}^{(1)} = 0$.
This result can also be found by a calculation of the relevant Feynman
diagram, which is
\[ 
\Sigma_{F}^{(1)} = - \frac{1}{4}\bar\lambda \int \frac{d^4
k}{(2\pi)^4}G^R_0(k) = 0,
\]
because all the poles are on the same side of the real $k^0$ axis in
the complex $k^0$ plane (see e.g. also \cite{AuBe92}).  The result is
that the second vertex in (\ref{eqpotrot}) does not contribute to this
order.

The classical retarded self energy to one loop is given by \be
\Sigma_{R, \rm cl}^{(1)} = \half\lambda \int \frac{d^4
k}{(2\pi)^4}S_0(k) - \lambda m_1^2, \ee where $m_1^2$ is given by
(\ref{ct1}). Comparing the classical and the quantum results, we see
that the classical result can simply be obtained with the substitution
$F_0 \to S_0$.  
Notice that to first order $\bar\lm$ and $\lm$ are
equal, as can be seen in the matching relation (\ref{eqmatchl}). The
difference appears in higher order. 
The final answer is
\[ \Sigma_{R, \rm cl}^{(1)} = -\lambda\frac{mT}{8\pi} - 
\lambda\frac{m^2}{32\pi^2} \log \frac{\bar\mu^2}{\mu^2_T}.
\]

The leading ${\cal O}(T)$ term (after resummation in the quantum
theory) is correctly reproduced by the classical one loop expression.
Indeed, as can be seen by explicitly writing $\hbar$, the term
proportional to $T$ is classical. 
The term with the logarithm, coming from the matching condition 
(\ref{ct1}), is proportional to $\hbar$.
We end the discussion of the one loop self energy with the remark that 
since this one loop diagram is momentum independent, the fact that we are
studying the {\em time dependent} two point function does not come
into play, and the outcome is the standard dimensional reduction
result.

We now turn to the second order contributions.

\subsubsection*{Second order retarded self energy: the setting sun diagrams}
The second order contribution to the self energy is more interesting,
because of explicit momentum dependence of the setting sun diagrams.

\begin{figure}
\centerline{\psfig{figure=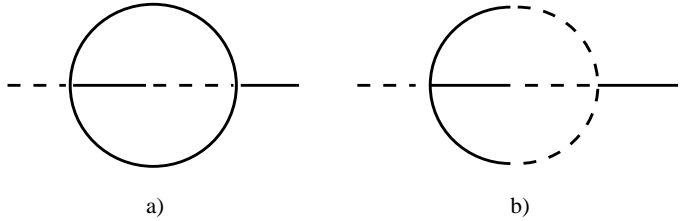,width=9cm}}
\caption{Setting sun contribution to the retarded self energy
$\Sigma_{R}^{\rm sun}$; in the quantum case both diagrams
contribute, in the classical case only the first one is present.}
\label{figsettingsun} 
\end{figure}

In the quantum theory the retarded setting sun diagrams are given
by\footnote{Diagrams that contain a subdiagram such as
\[ \int \frac{d^4k}{(2\pi)^4} G^R_0(k)G^R_0(p+k),\]
are not explicitly written, because they are zero (all the poles are
on the same side of the real $k^0$ axis in the complex $k^0$ plane).}

\bea
\label{eqsetsunqm1}
\Sigma_{R}^{\rm sun}(p) &=& -\half \bar\lambda^2\int
\frac{d^4k_1}{(2\pi)^4}\frac{d^4k_2}{(2\pi)^4} \Big[
F_0(k_1)F_0(k_2)G^R_0(p-k_1-k_2)\\ &&
\label{eqsetsunqm2} 
-\frac{1}{12} G_0^R(k_1)G_0^R(k_2)G_0^R(p-k_1-k_2)\Big].  \eea The
diagrams are given in fig \ref{figsettingsun}.  Note there is no
explicit temperature dependence in the second diagram.

After performing the temporal momentum integrals the first diagram can
be expressed as \be
\label{eqsetsun1}
\Sigma_{R, \rm a}^{\rm sun}(p) = -\frac{1}{2}\bar\lambda^2
\sum_{\{\ep_i\}} \int d\Phi_{123}(\vecp)\,
(n_{\veck_1}+\half)(n_{\veck_2}+\half)\frac{\ep_3}{p^0+i\ep+\ep_i\om_{\veck_i}},
\ee where $d\Ph_{123}(\vecp)$ was introduced in (\ref{eqdPh}),
$n_\veck = n(\om_\veck)$ and
\[ 
\ep_i\om_{\veck_i} = \ep_1\om_{\veck_1}+
\ep_2\om_{\veck_2}+\ep_3\om_{\veck_3}.
\]
The sum is over $\ep_i = \pm 1$, with $ i=1,2,3$.  The expression can
be written in a symmetric way by a permutation of the (1,2,3) indices.

The second diagram can be expressed as \be
\label{eqsetsun2}
\Sigma_{R, \rm b}^{\rm sun}(p) = -
\frac{\bar\lambda^2}{24}\sum_{\{\ep_i\}} \int d\Phi_{123}(\vecp)\,
\frac{\ep_1\ep_2\ep_3}{p^0+ i\ep +\ep_i\om_{\veck_i}}.  \ee One might
be surprised that we find two diagrams that contribute to the retarded
setting sun self energy. However, it is easy to check that the sum of
(\ref{eqsetsun1}) and (\ref{eqsetsun2}) gives precisely the expression
that is obtained in the imaginary time formalism after an analytic
continuation to real external frequencies \cite{WaHe96}.

The real part of the self energy contains the standard $T=0$
divergence, and a momentum independent subdivergence 
\cite{BaMa91,Pa92}. It satisfies 
$\mbox{Re}\; \Sigma_{R}^{\rm sun}(\vecp, -p^0) = \mbox{Re}\; 
\Sigma_{R}^{\rm sun}(\vecp, p^0)$. The imaginary part obeys 
$\mbox{Im}\; \Sigma_{R}^{\rm sun}(\vecp, p^0) 
= -\mbox{Im}\; \Sigma_{R}^{\rm sun}(\vecp, -p^0)$, so that we can
restrict ourselves to $p^0 > 0$. It can be written as \cite{WaHe96}
\[ 
- \mbox{Im}\; \Sigma_{R}^{\rm sun}(p) = g_1(p) + g_2(p),
\]
with 
\bean 
g_1(p) &=& \frac{\bar\lambda^2}{12}(e^{p^0/T}-1)\int d\Phi_{123}(\vecp)\, 
2\pi \delta(p^0 - \om_{\veck_1}-\om_{\veck_2}-\om_{\veck_3})
n_{\veck_1}n_{\veck_2}n_{\veck_3} ,\\ 
g_2(p) &=&
\frac{\bar\lambda^2}{4}(e^{p^0/T}-1) \int d\Phi_{123}(\vecp)\, 
2\pi \delta(p^0 + \om_{\veck_1}-\om_{\veck_2}-\om_{\veck_3})
(1+n_{\veck_1})n_{\veck_2}n_{\veck_3} , 
\eean 
representing three body
decay and Landau damping respectively.  The on-shell plasmon damping
rate 
\be 
\gm = \frac{-\mbox{Im}\;
\Sigma^{\rm sun}_{R} (\vecnul, m)}{2m} = \frac{g_2(\vecnul, m)}{2m} =
\frac{\bar\lambda^{3/2}T}{128\sqrt{6}\pi}
\left(1+{\cal O}(\sqrt{\bar\lambda}\log\bar\lambda)\right),
\label{qdamprate}
\ee
as calculated in \cite{Pa92, WaHe96}. 
The last equality is valid in the case that the thermal mass is much 
larger than the $\overline{\mbox{MS}}$ mass. 

In the classical theory we find the following expression 
\be
\Sigma_{R, \rm cl}^{\rm sun}(p) = -\half \lambda^2
\int\frac{d^4k_1}{(2\pi)^4}\frac{d^4k_2}{(2\pi)^4}
S_0(k_1)S_0(k_2)G^R_0(p-k_1-k_2) - \lambda^2 m_2^2, 
\ee 
with $m_2^2$
given by (\ref{ct2}).  It is similar to the quantum expression
(\ref{eqsetsunqm1}) after the replacement $S_0 \to F_0$
($\bar\lm$ and $\lm$ are the same to this order).
After this replacement we  see that, 
concerning the explicit temperature dependence,
(\ref{eqsetsunqm1}) is 
${\cal O}(T^2)$ and (\ref{eqsetsunqm2}) is ${\cal O}(T^0)$. Indeed, in the 
classical theory, which is supposed to be valid only to leading order in 
$T$,  there is no analogue of  (\ref{eqsetsunqm2}). 
Performing the temporal momentum integrals, we get
\bea
\label{eqsetsuncl}
\nonumber
\Sigma_{R, \rm cl}^{\rm sun}(p) &=&  
-\half\lm^2 T^2 \sum_{\{\ep_i\}} \int d\Phi_{123}(\vecp)\,
\frac{1}{\om_{\veck_1} \om_{\veck_2} \om_{\veck_3}}\,
\frac{\ep_3\om_{\veck_3}}{p^0+i\ep +\ep_i\om_{\veck_i}} - \lm^2m_2^2\\
\label{disprel}
&=& 
-\frac{\lambda^2T^2}{6} \int d\Phi_{123}(\vecp)\,
\frac{8}{\om_{\veck_1} \om_{\veck_2} \om_{\veck_3}}\,
+\frac{p^0}{T}\int \frac{d\Om}{2\pi}\frac{w(\vecp,\Om)}{p^0+i\ep +\Om} 
\\
\nonumber
&& -\lm^2m_2^2,
\eea
with
\[ w(\vecp, \Om) = 
\frac{\lambda^2}{6} 
\sum_{\{\ep_i\}}\int d\Phi_{123}(\vecp)\,
2\pi\delta(\Om - \ep_i\om_{\veck_i})
\frac{T^3}{\om_{\veck_1} \om_{\veck_2} \om_{\veck_3}}\,
.\]
In obtaining this result we symmetrized
$\ep_3\om_{\veck_3} \to \ep_i\om_{\veck_i}/3$. 
Notice that the first term in (\ref{disprel}) is real, 
independent of $p^0$ and logarithmically divergent.
This divergence is 
canceled with the counterterm $m_2^2$, cf.\ (\ref{statsetsun}) and
(\ref{ct2}). 
The weight function $w(\vecp,\Om)$ is finite and behaves like $\Om^{-1}$
as $\Om \to \infty$. Hence, the second term given by the dispersion
relation in (\ref{disprel}) is finite. 
So we arrive at the conclusion already obtained in 
\cite{AaSm97}, namely the classical theory can be made finite with just the 
counterterms of the static theory.
Furthermore, as we show in appendix \ref{appendixcomparison},
the difference between the quantum and classical
expressions for the two loop selfenergy 
is subleading for $T\to\infty$ and $\bar\lm \to 0$.

It is interesting to compare the imaginary parts of the selfenergies. 
Similar 
to the quantum case we can write the imaginary part of the classical
selfenergy as ($p^0 > 0$) 
\[ -\mbox{Im}\; \Sigma_{R, \rm cl}^{\rm sun}(p) = g_{1,\rm cl}(p) + 
g_{2, \rm cl}(p),\] with 
\bean 
g_{1, \rm cl}(p) &=& \frac{\lambda^2p^0}{12T}\int d\Phi_{123}(\vecp)\,
2\pi \delta(p^0 - \om_{\veck_1}-\om_{\veck_2}-\om_{\veck_3})
\frac{T^3}{\om_{\veck_1} \om_{\veck_2} \om_{\veck_3}}\,
,\\
g_{2, \rm cl}(p) &=& \frac{\lambda^2p^0}{4T} \int d\Phi_{123}(\vecp)\,
2\pi\delta(p^0 + \om_{\veck_1}-\om_{\veck_2}-\om_{\veck_3})
\frac{T^3}{\om_{\veck_1} \om_{\veck_2} \om_{\veck_3}}\,
.
\eean
Comparing the prefactors in these expressions with the quantum counterparts,  
we see explicitly that 
we must restrict ourselves to  external frequencies $|p^0| \ll T$
in order for the classical approximation to be valid. 
The classical on-shell damping rate is given by
\[ \gm_{\rm cl} = -\frac{\mbox{Im}\; \Sigma^{\rm sun}_{R, \rm cl} (\vecnul, 
m)}{2m} = \frac{g_{2, \rm cl}(\vecnul, m)}{2m} = \frac{\lambda^2 T^2}{1536 
\pi m} \approx \frac{\lambda^{3/2}T}{128\sqrt{6}\pi},\] 
where we used the matching relation (\ref{eqmatchm}) to obtain the last 
expression. 
After matching it indeed equals the quantum damping rate (\ref{qdamprate}), 
as we have shown earlier in \cite{AaSm97}.

Introducing $\hbar$ leads to the observation that the 
damping rate is independent of $\hbar$, if the classical mass parameter 
is arbitrary. However, 
taking the classical mass parameter according to the matching 
relation (\ref{eqmatchm}), gives a
damping rate which is proportional to $\hbar^{1/2}$ (when the thermal 
mass is much larger than the $\overline{\mbox{MS}}$ mass). 
This $\hbar$ dependence arises solely from the 
$\hbar$ dependence of the thermal mass.

Finally, the comparison between the quantum and the classical 
expressions in appendix \ref{appendixcomparison}, shows that the $p^0$ 
dependence is the 
same in leading order. Therefore we conclude that the possible prefactor 
$z_\pi$ in the kinetic part $\int d^3 x\, z_{\pi} \pi^2/2$ of the 
hamiltonian, is equal to one to this order.

\subsubsection*{Second order retarded self energy: the tadpole diagrams}

The tadpole diagrams are less interesting than the setting sun diagrams. 
Because they are momentum independent, the fact that we are concerned 
with time dependent correlation functions does not play a role. The 
matching procedure is identical to
 dimensional reduction
for static quantities. Therefore we will just give the expressions for 
completeness.    

\begin{figure}
\centerline{\psfig{figure=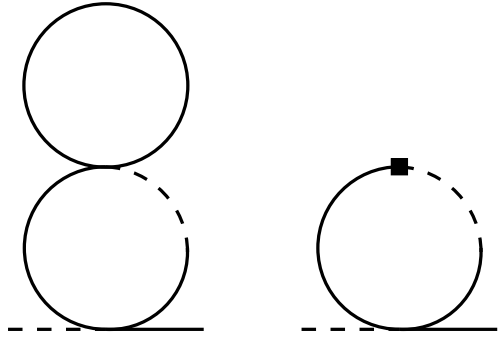,width=6.6cm}}
\caption{Tadpole contributions to the retarded self energy 
$\Sigma_{R}^{(2)\rm tp}$; the one loop diagram, with the vertex due to 
resumming, is only present in the quantum case.} 
\label{figtwotadpole}
\end{figure}

In the quantum theory we find the two loop and the one loop (with the vertex due to 
resumming) expressions 
\[
\label{eqqmpi12leaf}
\Sigma_{R}^{(2)\rm tp} = -\frac{1}{2} \bar\lambda^2
\int \frac{d^4k}{(2\pi)^4} F_0(k)G^R_0(k) \int \frac{d^4k'}{(2\pi)^4} 
F_0(k') 
+  \bar\lambda m_{\rm th}^2 \int \frac{d^4k}{(2\pi)^4} 
F_0(k)G^R_0(k), \]
which are shown  in fig \ref{figtwotadpole}. 
We have not shown the diagrams with $\overline{\mbox{MS}}$ 
counterterms insertions \cite{BaMa91, Pa92}.
Of course the tadpole contribution is real.

In the classical theory, we find the following second order contribution
\[ 
\Sigma_{R, \rm cl}^{(2)\rm tp} = -\frac{1}{2} \lambda^2\int 
\frac{d^4k}{(2\pi)^4} 
S_0(k)G_0^R(k)\int \frac{d^4k'}{(2\pi)^4}S_0(k')
+  \lambda^2 m_1^2 \int \frac{d^4k}{(2\pi)^4} 
S_0(k)G^R_0(k).
\]
Again the correspondence with the quantum expression follows from 
$F_0 \to S_0$. 

\subsubsection*{Second order $F$ component of the self energy}
For completeness we also give the $F$ component of the selfenergy. A 
detailed calculation is not necessary because it is related to the 
retarded self energy through the KMS condition (\ref{eqpiF}). It is however again 
instructive to compare the quantum with the classical expressions.

In the quantum theory we find
\bea \label{eqqmpi111} 
\Sigma_{F}^{(2)}(p) &=& -\bar\lambda^2\int 
\frac{d^4k_1}{(2\pi)^4}\frac{d^4k_2}{(2\pi)^4}
\Big[ \frac{1}{6}F_0(k_1)F_0(k_2)F_0(p-k_1-k_2) \\
&&
\label{eqqmpi112}
-\frac{1}{8}G^R_0(k_1)G^R_0(k_2)
\left(F_0(k_1 + k_2 -p)+ F_0(k_1+k_2 +p)\right)\Big],
\eea
shown in fig \ref{figtwof}. As can be seen in appendix 
\ref{appendixcounter}, there is 
no counterterm for these diagrams, they are finite by themselves.
\begin{figure}
\centerline{\psfig{figure=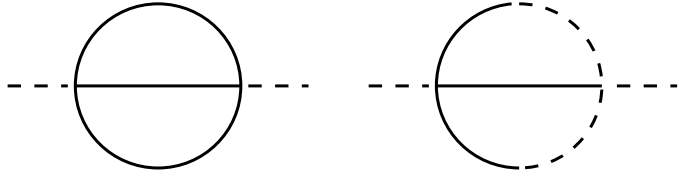,width=9cm}}
\caption{Setting sun contribution to the $F$ component of the self energy
$\Sigma_{F}^{(2)}$; in the classical case the second diagram is absent.}
\label{figtwof}
\end{figure}

In the classical theory we get
\[ \Sigma_{F, \rm cl}^{(2)}(p) = 
-\frac{\lambda^2}{6}\int \frac{d^4 
k_1}{(2\pi)^4} \frac{d^4 k_2}{(2\pi)^4} S_0(k_1)S_0(k_2)S_0(p-k_1-k_2).\]
Again we see that the classical expression is similar to 
the quantum expression (\ref{eqqmpi111}) after the replacement $F_0 \to S_0$. 
The second diagram (\ref{eqqmpi112}) has no classical counterpart, 
analogous to the situation for the retarded setting sun diagrams  
(\ref{eqsetsunqm1}) and (\ref{eqsetsunqm2}).

In the following section we take a look at the four point function.

\section{The four point function}
In this section we give the results for the four point vertex function. Since the static 
theory is superrenormalizable, the self coupling in the 
effective theory only receives a finite contribution due to matching (\ref{eqmatchl}). 
We show that also the time dependent four point function is finite.
Furthermore we show that 
the classical result gives the leading order quantum result.

The connected time dependent classical four point function is defined by 
\bean 
G_{\rm cl}(x_1, x_2, x_3, x_4) &=&
 \bra \phi(x_1)\phi(x_2)\phi(x_3)\phi(x_4)\ket_{\rm cl}\\ 
&&\;\;\;\;-\bra \phi(x_1)\phi(x_2)\ket_{\rm cl} \bra\phi(x_3)\phi(x_4)\ket_{\rm cl} + \mbox{perm.}
\eean
We extract the 1PI part in the usual way by an amputation of the 
external lines and use 
the obvious $1,2$ notation to indicate the full and dashed external lines. 
The external momenta are written as
$p \equiv p_1 +p_2 = p_3+p_4$. Diagrams that can simply be obtained 
by a permutation of the external lines are not explicitly written.
We also give the results for the 
zero loop vertex function, because this contains already non trivial 
information.

\begin{figure}
\centerline{\psfig{figure=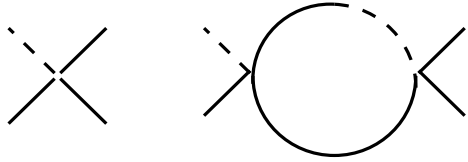,width=6.16cm}}
\caption{Zero and one loop contribution to the vertex function 
$\Gamma_{1112}$.}
\label{figfourp1222}
\end{figure}

We start with the $1112$ vertex function in the quantum theory. To second order 
in the coupling constant we find
\be 
\Gamma_{1112}^{(0+1)}(p) = -\frac{\bar\lambda}{3}
 + \bar\lambda^2\int \frac{d^4k}{(2\pi)^4}G^R_0(k+p)F_0(k) 
\ee
shown in fig \ref{figfourp1222}. 
The integral expression reads
\bea
\label{eqfourint}
\Gamma_{1112}^{(1)}(p) &=& \bar\lambda^2 \int
\frac{d^3k}{(2\pi)^3}\frac{1}{8\om_{\veck}\om_{\veck+\vecp}}
\Big[\\
&&
\nonumber (n_\veck+n_{\veck+\vecp}+1)
(\frac{1}{p^0+\om_\veck+\om_{\veck+\vecp}} - 
      \frac{1}{p^0-\om_\veck-\om_{\veck+\vecp}})\\
&&
\nonumber
+  (n_{\veck+\vecp}-n_\veck)
(\frac{1}{p^0+\om_\veck-\om_{\veck+\vecp}} - 
      \frac{1}{p^0-\om_\veck+\om_{\veck+\vecp}})\Big],
\eea
with $p^0 = p^0 + i\ep$. This quantum result contains the usual logarithmic 
$T=0$ divergence 
which is canceled with the counterterm (see appendix \ref{appendixcounter}). 
The leading order behaviour in the temperature is ${\cal O}(T)$.

In the classical case we find the, by now expected, result 
\be 
\Gamma_{1112, \rm cl}^{(0+1)}(p) = -\frac{\lambda}{3} 
+  \lambda^2\int \frac{d^4k}{(2\pi)^4}G^R_0(k+p)S_0(k) 
\ee
The integral expression is given by (\ref{eqfourint}) after the replacement 
$n_\veck + \half \to T/\om_\veck$. 
It is finite and it gives the leading ${\cal O}(T)$ quantum result. We 
discuss the difference between the quantum and the classical expressions  
in detail in appendix \ref{appendixcomparison}. 

For the case where the external spacial momenta $\vecp$ are zero, 
we have calculated the one loop contribution explicitly using standard 
high temperature techniques with the result
($y = m/T, s = p^0/2T$)
\bean
\bar\lambda^{-2}\Gamma^{(1)}_{1112}(\vecnul,p^0) &=& 
\frac{1}{16\pi}\frac{1}{\sqrt{y^2-s^2}+y} 
+\frac{1}{8\pi} \sum_{l=1}^{\infty} 
\frac{ \sqrt{y^2+(2\pi l)^2} -\sqrt{y^2-s^2} - 2\pi l}{(2\pi l)^2+s^2} \\
&&+
\frac{1}{16\pi^2}\sum_{n=1}^{\infty} (-1)^n
\left(\frac{s}{2\pi }\right)^{2n}\zt(2n+1)+ 
\frac{1}{32\pi^2}\left(2 +  \log \frac{\bar\mu^2}{\mu_T^2}\right),
\eean
with $\zeta(n)$ the Riemann zeta function. The classical result is simply
\[ 
\bar\lambda^{-2}\Gamma^{(1)}_{1112, \rm cl}(\vecnul, p^0) = 
\frac{1}{16\pi}\frac{1}{\sqrt{y^2-s^2}+y} 
+ \frac{1}{32\pi^2}\log\frac{\bar\mu^2}{\mu_T^2}
,\]
which is indeed the leading order quantum result when $m, p^0 \ll T$.
In obtaining this formula, we used the matching formula (\ref{eqmatchl}) 
to write $\lm$ in terms of $\bar \lm$. 

\begin{figure}
\centerline{\psfig{figure=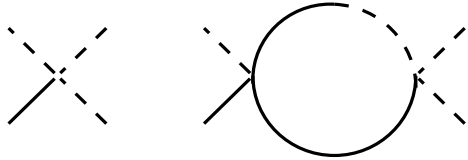,width=6.16cm}}
\caption{Zero and one loop contribution to the vertex function 
$\Gamma_{1222}$ in the quantum theory; in the classical theory it is zero.}
\label{figfourp1112}
\end{figure}

In the quantum theory we also have a 1222 vertex function. It turns out to be
\be 
\Gamma_{1222}^{(0+1)}(p) = \frac{1}{4}\Gamma_{1112}^{(0+1)}(p), 
\ee
and hence differs only in the external lines from $\Gamma_{1112}$, 
as shown in fig \ref{figfourp1112}. Adding the external lines to the 
1PI diagram to obtain the connected 
four point function, and using the fact that the external momenta 
$\vecp, p^0$ and the mass 
$m$ are small with respect to the temperature $T$, 
we find that the contribution from 
$\Gamma_{1222}$ to the connected four point function is ${\cal O}(T^{-2})$ 
suppressed with respect to the contribution from the $\Gamma_{1112}$ 
vertex function. Indeed, in the 
classical theory we do not have a 1222-type of vertex function.

\begin{figure}
\centerline{\psfig{figure=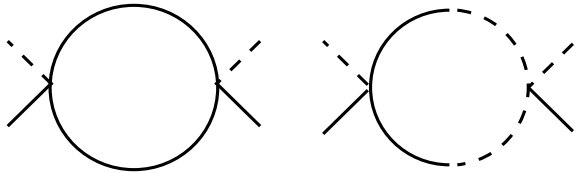,width=7.62cm}}
\caption{One loop contribution to the vertex function 
$\Gamma_{1212}$; in the classical case only the first diagram is present.}
\label{figfourp1212}
\end{figure}

For completeness we also give the result for the 1212 vertex function. 
It is related to the 1112 vertex function, as explained in 
appendix \ref{appendixvertex}. In the quantum theory we find
\be
\label{eqfourpoint1212}
\Gamma^{(1)}_{1212}(p) = 
\frac{\bar\lambda^2}{2}\int \frac{d^4k}{(2\pi)^4}\left( F_0(k)F_0(k+p) 
-\frac{1}{4}\left(G^R_0(k+p)+G^R_0(k-p)\right)G^A_0(k)\right),
\ee
there is no counterterm.  
The second diagram contains no explicit 
temperature dependence. In the classical theory we have \be
 \Gamma^{(1)}_{1212, \rm cl}(p) = \frac{\bar\lambda^2}{2}\int 
\frac{d^4k}{(2\pi)^4}S_0(k)S_0(k+p).\ee
The diagrams are shown in fig \ref{figfourp1212}.
Again the identification is straightforward, 
after the replacement $F_0 \to S_0$, the 
second quantum diagram is ${\cal O}(T^{-2})$ suppressed, 
and there is only one classical contribution.

This completes the discussion of the four point function. 
We conclude that the leading 
order quantum results are classical, when $\vecp, p^0, m \ll T$. 

\section{Conclusion}
In this paper we looked at real-time correlation functions in 
$\lambda\phi^4$ theory. We made a diagrammatic comparison between real-time 
correlation functions in quantum field 
theory and in the classical theory. The conclusions are twofold. 

The first conclusion is 
the one already obtained in \cite{AaSm97}, namely the classical theory is 
renormalizable.
A set of local counterterms is sufficient to cancel the
divergences that arise in the perturbative calculations. 
These counterterms are exactly the counterterms needed for the 
static correlation functions, which are dictated by the 
superrenormalizability of the static 
theory. This observation is 
useful for a non-perturbative calculation of time dependent classical 
correlation functions, e.g.\ 
numerically on a lattice. 
Lattice artefacts in physical observables
can be rendered negligible, in principle, 
by tuning of bare parameters.

The second conclusion is that at high temperature and small coupling
the classical theory can approximate the quantum theory. 
For this it is necessary to make a suitable choice of parameters in 
the effective classical hamiltonian, for which we used
standard dimensional reduction matching relations for time 
independent correlation functions. High temperature here means 
$\overline{\mbox{MS}}$ mass parameter $\bar m$, 
external momenta $\vecp$
and frequencies $p^0$ much smaller than $T$,
and $\overline{\mbox{MS}}$ schale $\bar\mu$ near
$\mu_T \approx 7T$. More precisely we found the condition
$\bar m, \vecp, |p^0| \leq {\cal O}(\sqrt{\bar\lm} T)$, and $\bar \lm 
\ll 1$.

An extension of these methods to gauge theories meets the problem 
that the perturbative infinities of the hot classical theory are 
similar in form to the hard thermal loop action \cite{BrPi90, TaWo90}, hence 
non-local (and on a lattice even non-rotational invariant) 
\cite{BoMcLeSm95, Ar97}. Yet, it may still be advantageous to view 
these non-local infinities as to be canceled by corresponding 
non-local counterterms, such that in principle the regularization 
can be removed. The ambiguity in the finite 
parts left after cancelation of the divergencies can then be removed
by matching with the standard hard thermal loop action. Such a 
procedure may be able to avoid the problem of regularization 
dependence, which enters approaches in which the cutoff is kept 
finite, of order of the temperature \cite{BoMcLeSm95,HuMu96}. 
Partial matching has already been used in numerical simulations 
of the classical SU(2)-Higgs theory at finite temperature, 
in which the effective parameters of the 
Higgs field were determined via dimensional 
reduction \cite{TaSm96,TaSm97,MoTu}. The hard thermal loop effects
of the gauge field were not matched, however.

\subsubsection*{Acknowledgements}

This work is supported by FOM.

\renewcommand{\thesection}{\Alph{section}}
\setcounter{section}{0}
\renewcommand{\theequation}{\Alph{section}.\arabic{equation}}

\section{Counterterms for the quantum theory}
\setcounter{equation}{0}
\label{appendixcounter}
We assume the quantum theory to be dimensionally renormalized 
in $d = 4 - 2\ep$ dimensions according to the $\overline{\mbox{MS}}$ scheme. 
The relation between $\mu$ of the MS scheme and $\bar{\mu}$ of the 
$\overline{\mbox{MS}}$ scheme is
\[ \mu = \bar{\mu}(4\pi e^{-\gm})^{-\half} 
\]
The counterterms ${\rm ct}$ in the actions (\ref{eqeucl}, \ref{eqaction})
are given by the coefficients $A$ -- $C$ below,
taken from ref \cite{BaMa91},
\bean 
{\rm ct} &=& 
\half\,  A \left( (\partial_\mu\phi_+)^2 - (\partial_\mu\phi_-)^2 \right) 
+ \half\, B \bar m^2 \left(\phi_+^2-\phi_-^2\right)
+\frac{1}{4!}\, C \bar\mu^{2\ep} \bar\lm\left(\phi_+^4-\phi_-^4\right)\\
&=&  
 A \partial_\mu\phi_1\partial^\mu\phi_2  
+ B \bar m^2 \phi_1\phi_2
+\frac{\bar \lambda}{4!}\, C \bar\mu^{2\ep}
\left(4\phi_1\phi_2^3 + \phi_1^3\phi_2\right).
\eean
The actual values of the counterterms are, 
up to the desired order and in the MS scheme 
\bean
A &=& \bar\lambda^2 A_2 = -\hat{\lambda}^2/24\ep,
\;\;\;\; \hat{\lambda} = \bar\lambda/16\pi^2,\\
B &=& \bar\lambda B_1 + \bar\lambda^2 B_2 = \hat{\lambda}/2\ep + 
\hat{\lambda}^2(1/2\ep^2-1/4\ep),\\
C &=& \bar\lambda C_1 = 3\hat{\lambda}/2\ep. 
\eean

\section{The classical KMS condition}
\setcounter{equation}{0}
\label{appendixKMS}
The classical KMS condition can be derived in the following way 
(see e.g \cite{Pa}).

Consider classical observables $A$ and $B$, obeying the equations of motion 
$\dot A = \{A, H\}$, $\dot B = \{B, H\}$, and look at the following 
correlation function
\[ \bra A(x)B(x')\ket_{\rm cl} = \frac{1}{Z} \int {\cal D}\pi {\cal D}\phi\, 
e^{-\beta H(\pi,\phi)} A(x)B(x').\]
Taking the derivative with respect to $t$ and using the equations of 
motion and the fact that the equations of motion and the canonical average 
are governed by the same hamiltonian, easily leads to
\bean
\frac{d}{dt} \bra A(x)B(x')\ket_{\rm cl} &=& 
\bra \{A(x),H\}B(x')\ket_{\rm cl}\\
&=& \frac{1}{\beta}\bra \{A(x),B(x')\}\ket_{\rm cl}.
\eean
Taking for both $A$ and $B$ $\phi$ and using the definition for the 
classical spectral function 
\[ \rho_{cl}(x-x') = -\bra \{\phi(x),\phi(x')\}\ket_{\rm cl},\]
gives the classical KMS condition
\[ \beta \frac{d}{dt}S(x-x') = -\rho_{cl}(x-x'),\]
or in temporal momentum space
\[
S(k) =  -\frac{T}{k^0}i\rho_{cl}(k).
\]

\section{The classical `mass derivative formula'}
\setcounter{equation}{0}
\label{appendixmassderivative}
In thermal field theory the doubling of the fields, the vertical part of the 
contour, the $i\ep$ regularization and the KMS condition, are all closely related. 
This can be checked in explicit calculations, i.e. with the so-called mass 
derivative formula \cite{LeB, Nie89}. We will repeat this calculation for the 
classical theory.

Consider the following hamiltonian
\bean
H &=& \int d^3x \left(\half \pi^2 + \half (\nabla \phi)^2 + \half m^2
\phi^2 + \half \kappa\phi^2 \right)\\
&=& H_0 +\int d^3x \half \kappa\phi^2.
\eean
The two point function, as defined in (\ref{eqdef2punt}),  is exactly 
given by
\[ 
S(\veck, t_1 -t_2) = T\frac{\cos [\sqrt{\om_\veck^2+ \kappa} (t_1-t_2)]}
{\om_\veck^2 +\kappa}, \]
with $\om_\veck^2 = \veck^2 + m^2$. We will, however, interpret the $\kappa$ term as an 
interaction vertex and 
calculate the  first order correction in $\kappa$ to the two point function. To 
check the result we can compare it with the exact result after expansion
\bea
\nonumber
S(\veck, t_1-t_2) &=& S_0(\veck, t_1-t_2) + \kappa S_1(\veck, t_1-t_2) + {\cal 
O}(\kappa^2), \\
\label{eqs1kappa}
S_1(\veck, t_1-t_2) &=& -\frac{T}{2\om_\veck^4} \left(\om_\veck (t_1-t_2) \sin 
\om_\veck (t_1-t_2) + 2\cos \om_\veck (t_1-t_2) \right).
\eea
We now  obtain this result by  solving the equations of motion perturbatively as
\[ \phi(\veck, t) =  \phi_0(\veck, t) + \kappa \phi_1(\veck, t) +{\cal 
O}(\kappa^2),\]
with the initial conditions
\[
\phi_0(\veck, t_0) = \phi(\veck), \;\;\;\;\dot\phi_0(\veck, t_0) = 
\pi(\veck),\;\;\;\; \phi_1(\veck, t_0) = \dot\phi_1(\veck, t_0) = 0.\]
$\phi_0(\veck, t)$ is given by (\ref{eqfree}), and
\[ \phi_1(\veck, t) =  \int_{t_0}^\infty dt'\, G_0^R(\veck, 
t-t')\phi_0(\veck, t').\]
We will now calculate the first order correction in two ways. 

\begin{enumerate}
\item[1.]
Keeping $t_0$ finite.
\end{enumerate}
We keep $t_0$ finite and expand also the Boltzmann weight as
\[ e^{-\beta H} = e^{-\beta H_0}\left(1-\beta \int d^3 x\, \half 
\kappa\phi_0^2(\vecx, t_0)\right).\]
$S_1(\veck, t_1-t_2)$ is found by collecting the two contributions proportional to 
$\kappa$. The result is
\bea
\label{eqexpandb}
S_1(\veck, t_1-t_2) &=& -\beta S_0(\veck, t_1-t_0)S_0(\veck, t_0-t_2) \\
&& 
\label{eqexpandeqm}
+\int_{t_0}^\infty dt'\, G_0^R(\veck, t_1-t')S_0(\veck, t'-t_2) + 
(t_1\leftrightarrow t_2).
\eea
(\ref{eqexpandb}) is coming from the first order expansion of the Boltmann weigth 
(proportional to $\beta$) and (\ref{eqexpandeqm}) from the first order solution to the 
equations of motion. Performing the integral gives
\bean
S_1(\veck, t_1-t_2) &=& 
-\frac{T}{\om_\veck^4}\cos \om_\veck(t_1-t_0)\cos \om_\veck(t_0-t_2)\\
&&-\frac{T}{\om_\veck^3} \int_{t_0}^{t_1} dt' \sin \om_\veck(t_1-t')\cos 
\om_\veck(t'-t_2) + (t_1\leftrightarrow t_2)\\
&=& 
-\frac{T}{2\om_\veck^4} \left(\om_\veck (t_1-t_2) \sin 
\om_\veck (t_1-t_2) + 2\cos \om_\veck (t_1-t_2) \right).
\eean
As expected the $t_0$ dependence has dropped out.
We stress that taking into account the terms coming from the Boltmann 
weight is crucial for this. We recover the result (\ref{eqs1kappa}).

\begin{enumerate}
\item[2.] Thermal field theory approach.
\end{enumerate}
In the standard thermal field theory approach $t_0$ is sent to $-\infty$ and 
contributions from the Boltmann weight (i.e. (\ref{eqexpandb})) are omitted. The part 
coming from the equations of motion (i.e. (\ref{eqexpandeqm})) is written 
in temporal momentum space. We get
\[ S_1(k) = \left( G_0^R(k) +G_0^A(k)\right)S_0(k),\]
with $S_0$ determined by the KMS condition
\be
\label{eqS} S_0(k) = -i\frac{T}{k^0}\left(G_0^R(k) - G_0^A(k)\right).\ee
This leads to 
\bean
S_1(\veck, t_1-t_2) &=& \int \frac{dk^0}{2\pi}\, e^{-ik^0(t_1-t_2)}S_1(k)\\
&=& \int \frac{dk^0}{2\pi i}\, e^{-ik^0(t_1-t_2)}\frac{T}{k^0}
\left({G_0^R}^2(k) - {G_0^A}^2(k)\right).
\eean
Closing the contour in the  lower (upper) halfplane for $t_1-t_2$ larger 
(smaller) than zero, and calculating the residues at the double poles at $k^0 = 
\pm \om_\veck - i\ep \;(\pm \om_\veck + i\ep)$ gives
\[ S_1(\veck, t_1-t_2) = -\frac{T}{2\om_\veck^4} \left(\om_\veck (t_1-t_2) \sin 
\om_\veck (t_1-t_2) + 2\cos \om_\veck (t_1-t_2) \right).\]
We remark that  in this case it is crucial that $S_0$ is given by (\ref{eqS}), 
in particular the $k^0$ dependence gives the correct contributions 
when calculating the residues at the poles.

We conclude that both methodes give the same, correct result. The standard 
thermal field theory approach can also be applied in the classical case.

\section{Comparison between quantum and classical  
diagrams: the vertex function and setting sun selfenergy}
\setcounter{equation}{0}
\label{appendixcomparison}
In this appendix we show that the classical diagrams (with the suitable 
matching relations) approximate the quantum expressions 
when the external momenta and frequencies are parametrically small
compared to the temperature, of order $\sqrt{\bar\lm}\, T$ for $\bar\lm \to 
0$. We will first discuss the vertex function, 
which is primitively divergent only at one loop order. 
It is illuminating to do the exercise, because 
it shows that the typical hard thermal loop structure is present, but not 
in leading order in the temperature. 
Then we discuss the setting sun contribution to  the retarded 
self energy, which is more involved.

The vertex function $\Gamma_{1112}^{(1)}(p)$ will be abbreviated to $\Gm(p)$ 
in  this appendix and we shall study 
$\Delta \Gm(p) \equiv \Gm(p) - \Gm_{\rm cl}(p)$.
To make the comparison easy, we introduce a momentum cutoff $\Lm$ to 
regulate the logarithmic divergence in the quantum diagram.
We add  a counterterm $C(\Lm)$ with the finite parts 
chosen in such a way that the renormalized vertex function is the same 
as in the $\overline{\mbox{MS}}$ scheme.

Using the result (\ref{eqfourint}) for the quantum and classical ($n_\veck 
+\half \to T/\om_\veck$) vertex function, we find for the difference
\bea
\nonumber
&&\bar\lm^{-2}\Delta \Gm(p) =  C(\Lm) - \frac{1}{32\pi^2}\log 
\frac{\bar\mu^2}{\mu_T^2} +  
\int \frac{d^3k}{(2\pi)^3}\frac{1}{8\om_{\veck}\om_{\veck+\vecp}}
\Big[\\
&&
\label{eqdiff1}
(n_\veck+n_{\veck+\vecp}+1 - \frac{T}{\om_\veck}
-\frac{T}{\om_{\veck+\vecp}} )
(\frac{1}{p^0+\om_\veck+\om_{\veck+\vecp}} - 
      \frac{1}{p^0-\om_\veck-\om_{\veck+\vecp}})\\
&&
\label{eqdiff2}
+ (n_{\veck+\vecp}-n_\veck - \frac{T}{\om_{\veck+\vecp}}
+\frac{T}{\om_{\veck}} )
(\frac{1}{p^0+\om_\veck-\om_{\veck+\vecp}} - 
      \frac{1}{p^0-\om_\veck+\om_{\veck+\vecp}})\Big],
\eea
The second term on the first line comes from the matching relation 
between $\lm$ and $\bar\lm$. 
It is convenient to rescale the variables as 
$\veck = \veck'T$, $\vecp = \sqrt{\bar\lm}\, \vecp'T$, 
$p^0 = \sqrt{\bar\lm}\, p'^0 T, \bar m = \sqrt{\bar\lm}\, \bar m' T$,
to indicate the magnitude of the momenta, frequencies and the 
$\overline{\mbox{MS}}$ mass parameter.

The first contribution (\ref{eqdiff1}) is the simplest. We take the 
limit $\bar\lm \to 0$, which gives an expression that is 
infrared finite, because the apparent 
infrared danger cancels between the classical and the quantum 
contribution. The result is independent of $p^{\mu}$.
Isolating the logarithmic divergence by adding and 
subtracting the integral
\[ \int_0^{\Lm/T} dk' \frac{1}{k'+1} = \log \Lm/T + {\cal O}(\Lm^{-1}),\]
we find for the difference ($k' = |\veck'|$)
\[
\bar\lm^{-2}\Delta \Gm(p)_{(\ref{eqdiff1})} =
 \frac{1}{8\pi^2} \int_0^{\infty}dk'
\left[ \frac{1}{k'}\left(\frac{1}{e^{k'}-1} +\half-\frac{1}{k'}\right) - 
\frac{1}{2(k'+1)}\right] + \frac{1}{16\pi^2}\log\frac{\Lm}{T}.
\]
The integral is infrared and ultraviolet finite. The log divergence is 
canceled with the counterterm $C(\Lm)$. 
The $\log T$ term is 
canceled by the matching relation,
the other terms are order $T^{0}$ and hence subdominant.

The second contribution (\ref{eqdiff2}) can be calculated by using standard 
HTL methods 
\cite{LeB}. Note that it is ultraviolet finite. 
We use
\bean 
(\om_{\veck+\vecp} - \om_{\veck})/T &=& \sqrt{\bar\lm}\, \hat \veck\cdot\vecp'
+ {\cal O}(\bar\lm), \\
n_{\veck+\vecp} - n_{\veck} &=& 
\frac{\partial n'_{\veck'}}{\partial k'}\,
\left[\om_{\veck+\vecp} - \om_{\veck}\right] + \cdots 
,\;\;\;\; n'_{\veck'} = (\exp k' -1)^{-1},
\eean
and similar for the difference of the classical distribution functions.
The angular integrals are now decoupled from the $k'$ integral, and the 
final result is ($p = |\vecp|$) 
\[
\bar\lm^{-2}\Delta \Gm(p)_{(\ref{eqdiff2})} =
\frac{1}{32\pi^2}\left( 2 + \frac{p^0}{p}\log \frac{p-p^0}{p+p^0}\right).
\]
This has the complicated momentum dependence of HTL expressions,
but it is order $T^0$, hence subdominant at high $T$.
{}From this we conclude that the leading order contribution in $T$ to the
quantum vertex function is given by the classical part.
The reason for this is the following.
If we leave out the vacuum part, and take the high temperature limit 
{\em under} the integral sign, (which means that we 
replace the Bose distribution with the classical one), the resulting 
integral is still ultraviolet finite.
Hence the integral is dominated by the low momenta, and the high 
temperature limit under the integral sign gives the leading order 
behaviour (this is of course not true if the resulting integral becomes 
divergent, like in e.g. the one loop tadpole diagram). 

Consider next the setting sun diagrams.
We start with a discussion of $\Sg_{R,\rm b}^{\rm 
sun}$, given in (\ref{eqsetsunqm2}). Classically, there 
is no diagram like this. 
Since all the internal lines are retarded Green 
functions, there is no explicit temperature dependence. Therefore, 
this diagram represents a part of the vacuum contribution. 
Indeed, the sum of the vacuum part of $\Sg_{R,\rm a}^{\rm sun}$ 
(given in (\ref{eqsetsunqm1})) with 
$\Sg_{R,\rm b}^{\rm sun}$ gives the normal $T=0$ expression.
One is inclined to think that the vacuum contribution is order $T^{0}$ 
and hence subdominant at high $T$. 
However, because of the resummation there is an implicit $T$ dependence. 
Because the diagram 
has dimension two, the momentum independent part is proportional to 
$\bar\lambda^2m^2$, which contains $\bar\lm^3 T^2$. So there is a $T^2$ 
contribution,  but fortunately it is suppressed
by one power of the coupling constant. Therefore the vacuum contribution 
is subdominant after all, 
when we restrict ourselves to a small coupling constant.

The classical contribution $\Sg_{R,\rm cl}^{\rm sun}$ is given in 
(\ref{eqsetsuncl}). A  remarkable 
property of the classical expression is that the logarithmic divergence 
and the $p^0$ dependence are completely separated. This is possible 
because of the symmetry of the integrand. 
It turns out that the quantum diagram can also be written in such a way, 
which makes the comparison easier.
After some rewriting, which is inspired by the Saclay method for calculating 
finite temperature diagrams, the setting sun contribution in 
the quantum theory can be written as
\bea
\nonumber
\Sg_{R}^{\rm sun}(p) &=& \Sg_{R,\rm a}^{\rm sun}(p) + \Sg_{R,\rm 
b}^{\rm sun}(p)\\
&=&
\nonumber
-\half\bar\lm^2\sum_{\{\ep_i\}}\int d\Phi_{123}(\vecp)
\frac{1-e^{-\beta\ep_i\om_{\veck_i}}}{p^0+i\ep+\ep_i\om_{\veck_i}}\,
\prod_{j=1}^3(n_{\veck_j}+\bar\ep_j)
\\
&=&
\label{eqcompare1}
-\half\bar\lm^2\sum_{\{\ep_i\}}\int d\Phi_{123}(\vecp)
\frac{1-e^{-\beta\ep_i\om_{\veck_i}}}{\ep_i\om_{\veck_i}}\,
\prod_{j=1}^3(n_{\veck_j}+\bar\ep_j)
\\
&&
\nonumber
+\half\bar\lm^2p^0\sum_{\{\ep_i\}}\int d\Phi_{123}(\vecp)
\frac{1}{p^0+i\ep+\ep_i\om_{\veck_i}}
\frac{1-e^{-\beta\ep_i\om_{\veck_i}}}{\ep_i\om_{\veck_i}}
\prod_{j=1}^3(n_{\veck_j}+\bar\ep_j),\\
&&
\label{eqcompare2}
\eea
where we used $\bar \ep_i = (\ep_i + 1)/2$.
Similar as in the classical diagram, there is no $p^0$ dependence in the 
first term (\ref{eqcompare1}). All the $p^0$ dependence (and hence the 
specific 
problems related to time dependence) is contained in (\ref{eqcompare2}). 
When we restrict ourselves to time independent 
correlation functions, only (\ref{eqcompare1}) is left over. Hence, this 
term is dealt by the dimensional reduction matching relations.
In particular, when the mass 
vanishes, it develops an infrared divergence, 
which causes it to behave like $\bar\lm^2 T^2\log m/T $ 
for $\bar\lm \to 0$ \cite{Pa92}. This term is matched (see (\ref{ct2})). 
Furthermore, the terms in the 
product that are linear in the Bose distribution give rise to a momentum 
independent logarithmic ultraviolet
subdivergence, proportional to $T^2$. This 
divergence cancels against a similar divergence in the second order 
tadpole contribution to the retarded self energy \cite{BaMa91}.

We now discuss the second term (\ref{eqcompare2}).
Writing out the product gives terms that have three, two, one and zero 
Bose distribution functions.  
Only the last (vacuum) 
contribution is ultraviolet divergent. 
There is no subdivergence.
(It would be proportional to $p^0$, but there is no such subdivergence
\cite{BaMa91}. The explicit $p^0$
factor cannot be canceled by the vanishing of $\ep_i\om_i$ in
the denominator $p^0 + i\ep + \ep_i\om_i$ because this can only cause a
$\log p^0$ behaviour, not a $1/p^0$ behaviour.)
Let's first look at the term with three Bose distributions. Taking the 
high temperature limit under the integral sign (as in the vertex function 
case), results in a ultraviolet finite expression. Hence this expression
gives the leading order 
behaviour. Furthermore we see that in this limit we recover in fact 
the second term in the classical expression (\ref{eqsetsuncl}). 
It now remains to show that 
the other terms (with two and one Bose distribution terms) are 
subdominant (here the high temperature limit cannot be taken under the 
integral sign!).  Rescaling the variables 
as in the vertex function case, gives a prefactor $\sqrt{\bar\lm}T^2$. 
Furthermore there are no infrared divergences for $\bar\lm \to 0$.
Hence, although these terms are proportional to $T^2$, they are 
suppressed by a power of $\sqrt{\bar\lm}$ 
(or $\sqrt{\bar\lm}\log \bar\lm$ 
in the case  $\ep_i\om_i$ vanishes in the denominator $p^0 + i\ep + 
\ep_i\om_i$).
For the real part of (\ref{eqcompare2}) we find a better result. Because 
it is even under $p^0 \to -p^0$, it is proportional to ${p^0}^2$, and 
hence after rescaling it is suppressed by a power of $\bar\lm$, instead 
of $\sqrt{\bar\lm}$.

The conclusion is 
that the leading part (at high $T$ and 
small $\bar\lm$) of the second term (\ref{eqcompare2}) is given by the 
corresponding classical expression.
 
The final  result of this appendix is that the classical vertex and 
setting sun diagrams give the leading 
order quantum result, after matching. This matching is identical to the 
conventional dimensional reduction matching for  time independent 
quantities.

\section{Relations between four point functions}
\setcounter{equation}{0}
\label{appendixvertex}
In this appendix we show that there is only one independent vertex 
function in $\phi^4$ theory, by use of the relation with $\phi^3$ theory. 

The potential in a $\phi^3$ theory becomes, after the transformation 
(\ref{eqrotphi}), 
\[  V(\phi) = \frac{g}{3!}\left(\phi_+^3 - \phi_-^3\right) = 
\half g\phi_1^2\phi_2 + \frac{g}{4!}\phi_2^3.
\]
The self energies are again related through the KMS condition
\[ \Sigma_{F}(p) = -i(n(p^0)+\half)\left(\Sigma_{R}(p) - \Sigma_{A}(p)\right).\]
The connection with the four point function in $\phi^4$ theory is straightforward
\[
\Sigma_{F}^{(1)}(p) = -\Gamma_{1212}^{(1)}(p),\;\;\;
\Sigma_{R}^{(1)}(p) = -\Gamma_{1112}^{(1)}(p).\]
Using these relations we find that only one vertex function  is 
independent in the $\phi^4$ theory.

\end{document}